\documentclass[letterpaper]{article}

\usepackage{geometry}
\usepackage{setspace}
\usepackage{graphicx}
\usepackage{float}
\newfloat{scheme}{htbp}{los}
\floatname{scheme}{Scheme}
\floatname{chart}{Chart}
\newfloat{graph}{htbp}{loh}

\usepackage{xurl}
\usepackage{authblk}
\usepackage{xspace}
\usepackage{float}
\usepackage{xspace}
\usepackage{graphicx}
\graphicspath{{./figures/}}
\usepackage{dcolumn}
\usepackage{bm}
\usepackage{color}
\usepackage{soul}
\usepackage{footnote}
\usepackage{multirow}
\usepackage{amsmath}
\usepackage{csquotes}
\usepackage{hyperref}
\hypersetup{
    pdftoolbar=true,        
    pdfmenubar=true,        
    pdffitwindow=false,     
    pdfstartview={FitH},    
    pdftitle={Fluctuation-enhanced-electron–phonon-coupling-in-FeSe},    
    pdfnewwindow=true,      
    colorlinks=true,       
    linkcolor=blue,          
    citecolor=blue,        
    filecolor=blue,      
    urlcolor=blue,       
    plainpages=false,        
}
\usepackage[english]{babel}
\usepackage{booktabs}
\usepackage{threeparttable}
\newcommand{\Ts}{\ensuremath{T_\mathrm{s}}\xspace}
\newcommand{\Tc}{\ensuremath{T_\mathrm{c}}\xspace}
\newcommand{\FESES}{\ensuremath{\mathrm{FeSe}_{1-x}\mathrm{S}_{x}}\xspace}

\newcommand{\grd}{$^{\circ}$\xspace}
\newcommand{\vei}{\ensuremath{\mathbf{e}_\mathrm{i}}\xspace}
\newcommand{\ves}{\ensuremath{\mathbf{e}_\mathrm{s}}\xspace}
\newcommand{\wn}{\ensuremath{\rm cm^{-1}}\xspace}
\newcommand{\Alg}{\texorpdfstring{\ensuremath{A_{1g}}\xspace}{A1g}}
\newcommand{\Algprime}{\texorpdfstring{\ensuremath{A_{1g}^{\prime}}\xspace}{A1g'}}
\newcommand{\Algph}{\texorpdfstring{\ensuremath{A_{1g}^{ph}}\xspace}{A1gph}}

\newcommand{\Blgph}{\texorpdfstring{\ensuremath{B_{1g}^{ph}}\xspace}{B1gph}}

\newcommand{\Blg}{\texorpdfstring{\ensuremath{B_{1g}}\xspace}{B1g}}
\newcommand{\BZg}{\texorpdfstring{\ensuremath{B_{2g}}\xspace}{B2g}}

\providecommand{\JournalTitle}[1]{#1}
\providecommand{\bibinfo}[2]{#2}
\DeclareRobustCommand{\onlinecite}[1]{\cite{#1}}

\author[1,$\dagger$]{Jovan Blagojevi\'{c}}
\author[1,$\dagger$, *]{Ana Milosavljevi\'{c}}
\author[1]{Tea Belojica}
\author[1]{Marko Opa\v{c}i\'{c}}
\author[1]{Andrijana \v{S}olaji\'{c}}
\author[1]{Jelena Pe\v{s}i\'{c}}
\author[2]{Enrico Di Lucente}
\author[1]{Novica Paunovi\'{c}}
\author[3,4]{Milorad V. Milo\v{s}evi\'{c}}
\author[1]{Emil S. Bozin}
\author[5]{Aifeng Wang}
\author[6,7,5,8]{Cedomir Petrovic}
\author[9]{Zoran V. Popovi\'{c}}
\author[10,11]{Rudi Hackl}
\author[1,12, *]{Bojana Vi\v{s}i\'{c}}
\author[1]{Nenad Lazarevi\'{c}}

\affil[1]{Center for Solid State Physics and New Materials, Institute of Physics Belgrade, University of Belgrade, Pregrevica 118, 11080 Belgrade, Serbia}
\affil[2]{Department of Applied Physics and Applied Mathematics, Columbia University, New York, USA}
\affil[3]{COMMIT, Department of Physics, University of Antwerp, Groenenborgerlaan 171, B-2020 Antwerp, Belgium}
\affil[4]{Stavropoulos Center for Complex Quantum Matter, Department of Physics and Astronomy, University of Notre Dame, Notre Dame, Indiana 46556, USA}
\affil[5]{Condensed Matter Physics and Materials Science Department, Brookhaven National Laboratory, Upton, New York 11973-5000, USA}
\affil[6]{Shanghai Advanced Research in Physical Sciences (SHARPS), Shanghai 201203, China}
\affil[7]{Center for High Pressure Science \& Technology Advanced Research (HPSTAR) - Beijing 100094, China}
\affil[8]{Department of Nuclear and Plasma Physics, Vinca Institute of Nuclear Sciences, University of Belgrade, Belgrade 11001, Serbia}
\affil[9]{Serbian Academy of Sciences and Arts, Kneza Mihaila 35, 11000 Belgrade, Serbia}
\affil[10]{School of Natural Sciences, Department of Physics E51, Technische Universit\"{a}t M\"{u}nchen, 85748 Garching, Germany}
\affil[11]{IFW Dresden, Helmholtzstrasse 20, 01069 Dresden, Germany}
\affil[12]{Department of Condensed Matter Physics, Jozef Stefan Institute, Jamova cesta 39, Ljubljana 1000, Slovenia}

\title{Fluctuation-enhanced electron–phonon coupling in FeSe}
\date{%
\textsuperscript{$\dagger$}These authors contributed equally\\
\textsuperscript{*}Corresponding authors email: ana.milosavljevic@ipb.ac.rs; bojana.visic@ijs.si%
}
\begin{document}

\maketitle

\begin{abstract}

The interactions among lattice, charge, and spin degrees of freedom fundamentally shape material properties. In FeSe, symmetry-breaking perturbations serve as highly sensitive probes of these couplings. Previous work has shown that defects and isoelectronic substitution can substantially alter these interactions, giving rise to additional phonon modes. In this study, uniaxial strain is employed as a tunable symmetry-breaking control parameter to probe the intrinsic lattice response in the absence of disorder. The temperature evolution of phonon excitations was examined with fine temperature resolution in the vicinity of the nemato-structural transition temperature \Ts, under strain applied along the $\langle110\rangle$ and $\langle100\rangle$ crystallographic directions. A subtle asymmetry of the \Algph mode appears in the unstrained crystal within a narrow temperature window around \Ts, originating from the emergence of an additional mode in the fully symmetric channel. With applied strain, this feature becomes more distinctly resolved. The anomaly is attributed to modifications of the coupling between lattice and electronic degrees of freedom driven by the ordering fluctuations right above the nematic transition. These fluctuations enhance susceptibility for phonon-electron-phonon coupling in the vicinity of the $X$ and $R$ points of the Brillouin zone and promote two-phonon scattering close to the \Algph mode. The presence of this two-phonon scattering depends on both the strength and the direction of the applied strain, indicating a high sensitivity of FeSe to local symmetry breaking.

\end{abstract}
\flushbottom

\thispagestyle{empty}
\section*{Keywords}
{Strain, defects, iron based superconductors}

\section*{Introduction}

FeSe is the simplest iron-based superconductor exhibiting \Tc = 9~K at ambient pressure.~\cite{FeSe_SC_Fong_2008} Superconductivity is substantially enhanced in bulk samples under pressure, with \Tc reaching 36~K at 10~GPa,~\cite{Massat2018jan, Margadonna2009_PRB80_064506}  while thin films display even higher \Tc values of about 50~K.~\cite{Qing-Yan2012_CPL3_37402, Tan2013_NM12_634} Substituting 17\% of Se with S raises \Tc to 11~K,~\cite{Matsuura2017_NC8_1143} while intercalating alkali ions or molecular layers between the FeSe sheets can boost it above 40~K.~\cite{Guobing_FeSeIntercal_2021} These results demonstrate that even small modifications of the structure can lead to substantial changes in the properties of FeSe. Consequently, since superconductivity is a highly sensitive indicator of interactions between electrons and the related excitations of the spin and lattice degrees of freedom, FeSe serves as an excellent platform for disentangling the various types of coupling mechanisms that remain under debate in iron-based systems. Among the different internal and external perturbations, the application of uniaxial stress specifically targets the interplay of the lattice with predominantly orbital degrees of freedom.

Angle-resolved photoemission spectroscopy (ARPES) has revealed that the low-energy electronic structure of FeSe consists of three hole-like bands around the $\Gamma$ point and two electron-like bands centered at $M$.~\cite{PhysRevB.94.201107, Watson_2017, Rhodes2022FeSe, PhysRevB.92.205117, PhysRevB.101.235128,  Coldea2018_ARCMP9_125} In the tetragonal phase, the two electron pockets at $M$ are nearly circular and symmetry-equivalent, arising from hybridized $d_{xz}$ and $d_{yz}$ orbitals. Below the nemato-structural transition this orbital degeneracy is lifted, producing pronounced anisotropic shifts and hybridization gaps near $M$. As a consequence, only one of the two electron bands continues to cross the Fermi level, while the other one is pushed above it, an effect commonly referred to as the "missing-electron-pocket problem".~\cite{Rhodes2022FeSe, PhysRevB.94.201107, PhysRevX.9.041049} In twinned crystals, where orthorhombic domains of both orientations coexist, the ARPES intensity maps at $M$ reveal two perpendicular, peanut- or propeller-shaped Fermi surface contours originating from the two domain variants. In detwinned crystals, where uniaxial strain aligns the domains to a single orthorhombic one, only one of these features remains visible, reflecting the intrinsic electronic reconstruction driven by nematic order. Simultaneously, the hole band at $\Gamma$ sinks below the Fermi level and becomes anisotropic, consistent with orbital polarization between $d_{xz}$ and $d_{yz}$ states below \Ts.~\cite{PhysRevB.94.201107, Watson_2017}

Building on these observations, an open question remains as to how the delicate balance between lattice and electronic degrees of freedom in FeSe responds to externally applied symmetry-breaking fields. The missing electron pocket at the $M$ point and the pronounced orbital polarization indicate that the electron–phonon coupling (EPC) in this material is inherently anisotropic and strongly tied to local symmetry.~\cite{PhysRevB.100.020501, FrandsenB_FeSe_Structure_2019} By selectively breaking the in-plane fourfold symmetry, strain field directly interacts with the nematic order parameter and tunes the lattice potential without introducing embedded disorder. Ideally, uniaxial strain would allow us to disentangle the role of electron-electron and electron-phonon coupling in FeSe.

To address this, we employ uniaxial strain as a clean and reversible control parameter, focusing on the narrow temperature range around the nemato-structural transition, where EPC effects are expected to be strongest. By tracking the temperature evolution of zone-center phonons under strain applied along the $\langle110\rangle$ and $\langle100\rangle$ crystallographic directions, we demonstrate that the direction and symmetry of the applied field dictate the lattice response and its coupling to electronic fluctuations. This analysis reveals that the asymmetry of the \Algph mode originates from the electronic reconstruction near the $M$ point and the associated modifications of EPC, analogous to those observed in FeS and FeSe$_{1-x}$S$_x$. The results show that the vibrational properties of FeSe are directly governed by its correlated electronic state, underscoring the mutually dependent evolution of lattice and electronic fluctuations near \Ts. Altogether, FeSe emerges as a system where symmetry breaking, nematic fluctuations, and EPC are inseparably entangled, such that even subtle perturbations to the lattice or electronic structure lead to disproportionately large and highly anisotropic responses.

\section*{Results and discussion}

\subsection*{Uniaxial strain, symmetry and selection rules}

\begin{figure}[!b]
  \centering
  \includegraphics[width=165mm]{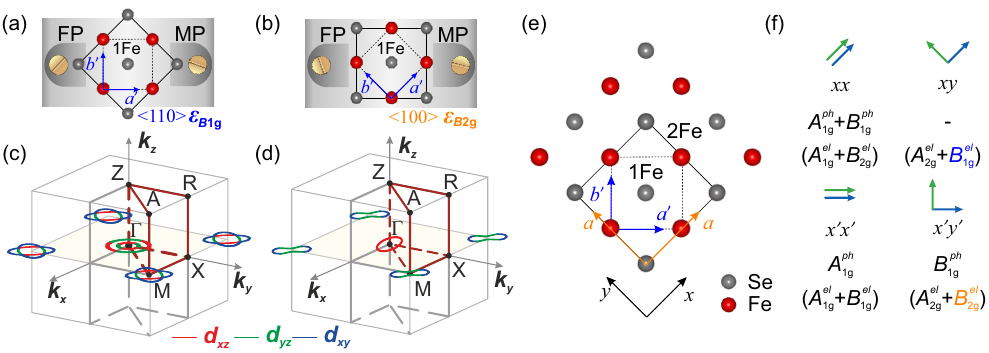}
  \caption{(a) and (b) Diagram showing the applied strain direction with respect to the 1-Fe unit cell. Fixed and movable sample plate integrated to piezo-electric device are labeled as FP and MP, respectively. (c) Schematic representation of the FeSe 2-Fe Brillouin zone and Fermi surface in tetragonal and (d) nematic phase $\left( \text{Adapted from Ref.~\onlinecite{Rhodes2022FeSe}} \right)$. (e) Crystal structure of FeSe in the $ab$ plane ($b=a$). The solid line represents the 2-Fe crystallographic unit cell determining the symmetry of the lattice vibrations. The dashed line represents the 1-Fe unit cell with axes $a'$ and $b'$ relevant for electronic and spin excitations. (f) Selection rules for four main linear polarization configurations. The light polarizations $x$ and $y$ are aligned with the crystallographic axes ($a,a$). The Raman-active \Algph and \Blgph phonons are projected in parallel $x'x'$ and perpendicular $x'y'$ scattering configurations, respectively. Electronic and spin excitations are preferably described in the 1-Fe cell, where \Blg and \BZg are swapped with respect to the $xy$ system (symmetries in brackets). These symmetries are used for the application of strain (blue and orange, respectively).}
 \label{fig:Figure1}
\end{figure}

The experimental geometry and the two strain configurations employed in this study are illustrated in Fig.~\ref{fig:Figure1} and described in the Methods section. In one configuration the $a'$ axis (aligned with Fe-Fe bond direction) is parallel to the strain direction [Fig.~\ref{fig:Figure1}~(a), \Blg, see Fig.~\ref{fig:Figure1}~(e) for definition]. In the other one the strain is applied along the crystallographic $a$ axis or at 45\grd with respect to the $a'$ direction [Fig.~\ref{fig:Figure1}~(b), \BZg, see Fig.~\ref{fig:Figure1}~(e) for definition]. These two geometries were selected as they impose distortions of different symmetry: (i) strain along $a'$  ($\epsilon_{\Blg}$) couples directly to the nematic \Blg distortion that describes the transition of the quadratic 1-Fe unit cell into an orthorhombic one; (ii) strain along $a$ transforms the 1-Fe tetragonal cell into a diamond or rhombohedron (\BZg symmetry; see Methods).  This approach enables a direct comparison of different symmetry-breaking mechanisms and their impact on the lattice dynamics.

Since \Blg strain directly couples to the nematic order parameter, it is important to recall how the underlying electronic structure evolves across the nemato-structural transition. The corresponding Fermi surface topology above and below \Ts is schematically illustrated in Fig.~\ref{fig:Figure1}~(c) and (d), respectively, for the 2-Fe cell. In the tetragonal phase [Fig.~\ref{fig:Figure1}~(c)], FeSe hosts two hole-like bands centered at $\Gamma$ and two symmetry-equivalent elliptical electron pockets at the $M$ point (the Brillouin zone of 1-Fe cell is twice as large and $M$ becomes $X$). Below \Ts [Fig.~\ref{fig:Figure1}~(d)], lifting of the $d_{xz}$/$d_{yz}$ orbital degeneracy leads to an anisotropic reconstruction of the electronic states~\cite{PhysRevB.94.201107, Watson_2017, Rhodes2022FeSe} and propellers oriented along ${\bf k}_x$ of the 1-Fe cell.

The selection rules governing phonon Raman scattering are dictated by the symmetry of the crystallographic 2-Fe unit cell. In our case, only in-plane polarizations (within the $ab$ plane) are applied, as illustrated in Fig.~\ref{fig:Figure1}~(e), where solid and dashed lines represent the 2-Fe and 1-Fe unit cell, respectively. While the 1-Fe unit cell is often used to describe electronic and spin excitations, since the relevant bands and the magnetism derive from the Fe orbitals, the 2-Fe unit cell provides the appropriate framework for analyzing phonon modes.

Incident and scattered photons propagate approximately along the $c$-axis. Under these conditions, the selection rules dictate that only \Algph and \Blgph Raman-active modes are observable for the $P4/nmm$ (space group 129) symmetry. Fig.~\ref{fig:Figure1}~(f) shows which phonon modes are projected for a given scattering geometry. In $x'x'$ (see Methods section) configuration, the scattering probes only the non degenerate fully symmetric in-phase vibration of Se atoms along the $c$-axis (\Algph) since there is no \BZg phonon. In the crossed $x'y'$ polarization, the observed \Blgph mode breaks the fourfold $C_4$ rotational symmetry and corresponds to the out-of-phase vibration of Fe atoms along the $c$-axis.

\subsection*{Raman spectra of FeSe}

\begin{figure*}[!b]
  \centering
  \includegraphics[width=165mm]{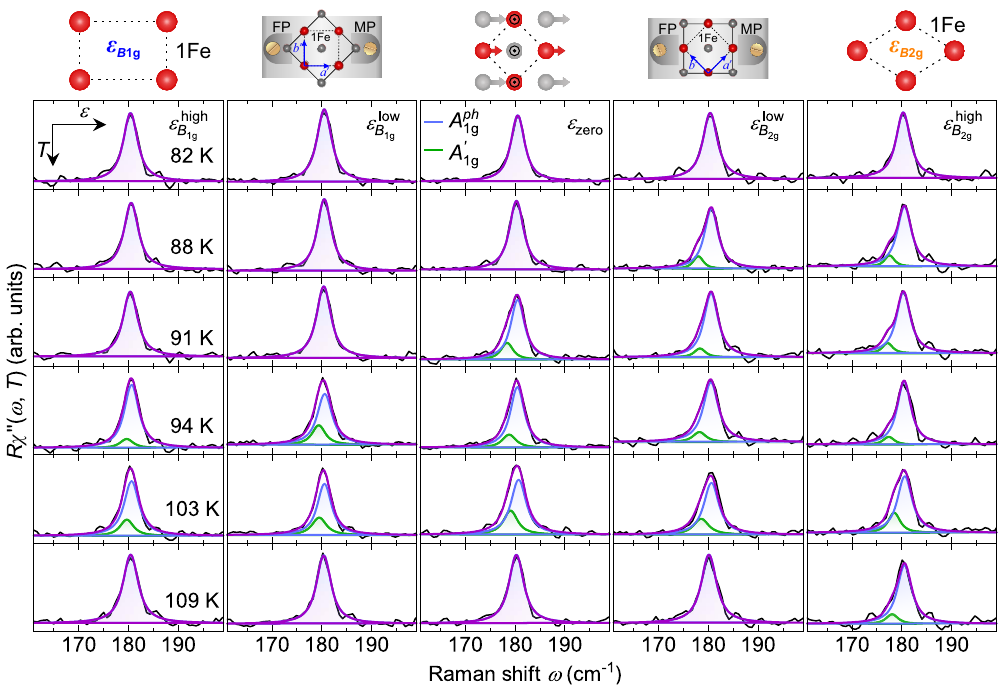}
  \caption{Top panel: Illustrations at left (right) schematically depict distortion of 1-Fe unit cell and sample orientation under corresponding strain $\varepsilon_{\Blg}$ $(\varepsilon_{\BZg})$, while the middle one represents atomic displacement of (acoustic) vibration at the X-point of the Brillouin zone (2-Fe crystallographic unit cell). Spectra panels: Strain-dependent Raman spectra in $x'x'$ scattering geometry at temperatures indicated, in the spectral range where the \Algph phonon is expected. Note that the $xy$ system of the polarizations is always aligned parallel to $ab$. Arrows in the legend indicate change in temperature and in strain configuration, i.e. $\varepsilon^{\mathrm{high}}_{\Blg}$, and $\varepsilon^{\mathrm{low}}_{\Blg}$ denote spectra obtained under uniaxial tensile strain applied along $a'$-axis, while $\varepsilon^{\mathrm{high}}_{\BZg}$ and $\varepsilon^{\mathrm{low}}_{\BZg}$ correspond to spectra obtained under tensile strain along the diagonal of 1-Fe unit cell. $\varepsilon_{\mathrm{zero}}$ labels the spectra recorded in the absence of applied strain. The blue-shaded area marks the symmetry-allowed \Algph mode, and the green-shaded area highlights an additional feature emerging near \Ts, labeled \Algprime. 
  }
	\label{fig:Figure2}
\end{figure*}
The nemato-structural phase transition in FeSe occurs at \Ts = 89.1\,K [see Fig.~S1 of Supplementary Information (SI)],~\cite{Baum2019_CP2_14, PhysRevB.100.020501, FrandsenB_FeSe_Structure_2019} with crystal symmetry decreasing from tetragonal $P4/nmm$ to orthorhombic $Cmmn$ symmetry. The transition leaves $b=a$, but introduces a small deviation from 90\grd between $a$ and $b$~\cite{doi:10.1126/sciadv.aar6419, Chubukov2015_PRB91_201105, PhysRevB.106.094510, PhysRevB.100.020501, Baek2015_NM14_210} of 0.23\grd, for $\delta = (a'-b')/(a'+b') \approx 0.002$.~\cite{Boehmer2018_JPCM30_023001, PhysRevB.105.064505} Although this deviation does not affect the selection rules, there are subtle spectral changes across this transition and we performed detailed temperature-dependent Raman scattering measurements with temperature steps of 3\,K in the vicinity of \Ts.

We first examine the $x'x'$  spectra. Fig.~\ref{fig:Figure2} (spectra panels) shows the spectral region of the \Algph phonon as a function of temperature and applied tensile strain. From top to bottom we show six representative temperatures, from left to right we show results for different strain directions and magnitudes. Specifically, the first two columns correspond to \Blg, inducing an orthorhombic distortion of the Fe plane, while the last two columns correspond to \BZg strain (see top panel). The spectra in the middle column represent the unstrained case $\left(\varepsilon_{\mathrm{zero}}\right)$. This compilation enables a direct comparison of how distinct strain configurations affect the \Algph phonon evolution with temperature. From left to right, the effects of strain become progressively more evident, illustrating its systematic impact on the phonon response.

In the unstrained case, an additional peak (green shading) appears within the narrow range between 91 and 103\,K on the low-energy side of the \Algph phonon (blue shading). We label this new feature as $\Algprime$ to denote its fully symmetric character in the FeSe compound (see also SI). The mode appears when the long range orthorhombic distortion vanishes and survives in the temperature range where the critical fluctuations reach their maximum.~\cite{Baum2019_CP2_14}

A similar behavior is observed for \Blg strain  coupling directly to nematic order and enhancing the orthorhombic distortion of the Fe plane (Fig.~\ref{fig:Figure2}, left columns). The additional $\Algprime$ peak appears again, only at higher temperature (94 – 106\,K) compared to the unstrained case.

In contrast, \BZg strain distortion produces a considerably different evolution (Fig.~\ref{fig:Figure2}, right columns). For low-magnitude strain $(\varepsilon^{\mathrm{low}}_{\BZg})$, the $\Algprime$ feature is observed over an extended temperature interval ranging from 88 to 106\,K and appears as a well-defined shoulder on the low-energy side of the \Algph. With increasing strain $(\varepsilon^{\mathrm{high}}_{\BZg})$, the feature emerges at the same temperature as for $\varepsilon^{\mathrm{low}}_{\BZg}$ but remains visible up to about 109\,K, indicating a slight broadening of the anomaly window.

\begin{figure}[!b]
  \centering
  \includegraphics[width=85mm]{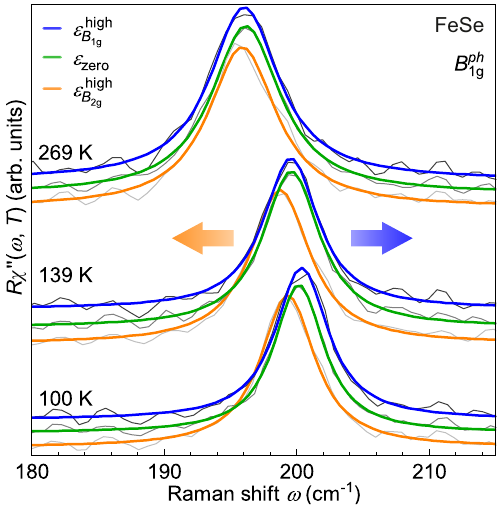}
  \caption{Strain-dependent \Blgph Raman spectra at temperatures as indicated. At 269~K strain has only very little effect. Upon approaching the transition at \Ts,  \Blg strain leads to a blue shift (blue) of the phonon, while \BZg strain causes a red shift (orange) with respect to the unstrained case (green).}
 \label{fig:Figure3}
\end{figure}

To become more quantitative we described all spectra using Voigt profiles (see Methods). None of the spectra in the temperature ranges indicated above could be described reasonably well using a single line. The discrepancy was most evident for \BZg strain (Fig.~\ref{fig:Figure2}, right columns), where the line shape directly indicates the presence of an additional peak on the low-energy side. When a second Voigt component was introduced, the fit quality improved significantly for both the unstrained and strained cases, within the corresponding temperature ranges discussed above. A comparison between single- and two-component fits is presented in Fig.~S2 of the SI.

We further analyze the response of the lattice to strain and temperature in the $x'y'$ scattering channel. Fig.~\ref{fig:Figure3} shows Raman spectra in the energy range of the \Blgph mode at selected temperatures. The  orthorhombic distortion ($\varepsilon_{\Blg}$) causes the \Blgph phonon to shift to slightly  higher energies at low temperatures, whereas rhombohedral distortion ($\varepsilon_{\BZg}$) results in shifts of about 1~\wn toward lower energies. As the temperature increases, these strain-induced shifts progressively decrease, and the phonon energies converge toward those of the unstrained sample.

\subsection*{Temperature dependence of the phonon excitations}

\begin{figure}[!ht]
  \centering
  \includegraphics[width=165mm]{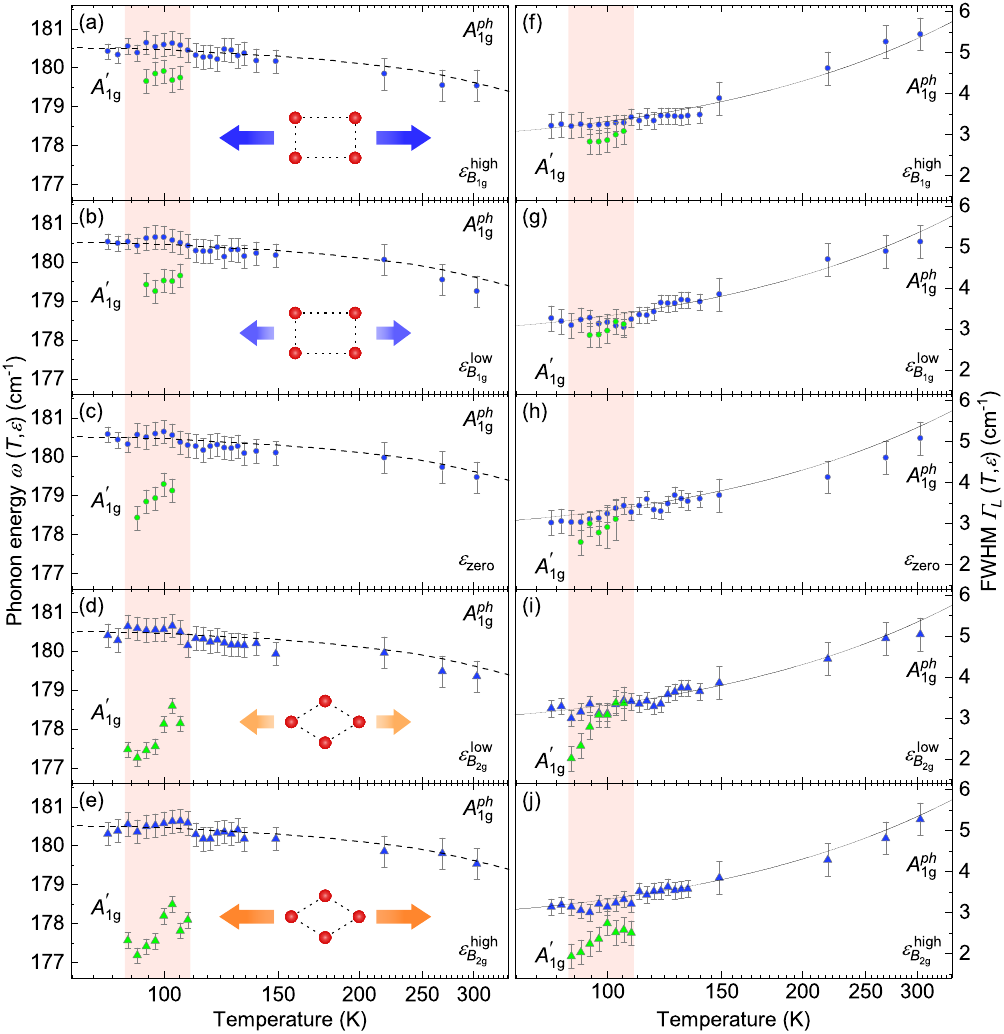}
  \caption{Temperature dependence of the \Algph and \Algprime mode energies (left panel) and Lorentzian linewidths (right panel) for different strain configurations. From top to bottom, panels show the phonon energy and linewidth evolution for (a,f) $\varepsilon_{\Blg}^{\mathrm{high}}$, (b,g) $\varepsilon_{\Blg}^{\mathrm{low}}$, (c,h) $\varepsilon_{\mathrm{zero}}$, (d,i) $\varepsilon_{\BZg}^{\mathrm{low}}$, and (e,j) $\varepsilon_{\BZg}^{\mathrm{high}}$. The black (a-e) dashed  and (f-i) solid lines represent the temperature dependencies of the phonon energies [according to Eq.~(\ref{eq:energy})] and linewidths [according to Eq.~(\ref{eq:linewidth})], respectively.}
 \label{fig:Figure4}
\end{figure}

Using the results of the quantitative analysis, we now focus on the temperature dependences of the energies and linewidths of all detected excitations in the zero-strain regime, as well as under low and high strain conditions along the two directions considered. The extracted phonon energies and Lorentzian linewidths of the \Algph and  \Algprime modes are presented in Fig.~\ref{fig:Figure4}, in the left and right panel respectively. The temperature dependence of the linewidths was further modeled using the standard expression for symmetric anharmonic decay~\cite{Klemens1966_PR148_845}:

\begin{equation}
\mathrm{\Gamma_L}(T,\varepsilon) = \mathrm{\Gamma_L(0)}\left(1+\frac{2\mathrm{\lambda_{ph-ph}}}{\mathrm{e}^{\frac{\hbar {\omega}(0)}{2\mathrm{k_B}T}}-1}\right),
\label{eq:linewidth}
\end{equation}

\noindent where $\mathrm{\Gamma_L(0)}$ and ${\omega}(0)$ were obtained by extrapolating linewidths and energies to the zero temperature limit, while  phonon-phonon coupling constant ($\mathrm{\lambda_{ph-ph}}$) is a free parameter. The phonon energy $\omega(T,\varepsilon)$ is influenced by both anharmonic decay and lattice contraction, and  is given by the following expression~\cite{Eiter2014_PRB90_024411}:

\begin{equation}
\begin{split}
\omega(T,\varepsilon) &= \omega(0)\left[1-\gamma \frac{V(T)-V(0)}{V(0)} - \frac{\Gamma^2_L(0)}{2\omega^2(0)} \left(1 + \frac{4\mathrm{\lambda_{\text{ph-ph}}}}{\mathrm{e}^{\frac{\hbar {\omega}(0)}{2 k_B T}}-1}\right) \right].
\end{split}
\label{eq:energy}
\end{equation}

\noindent Here, the volume $V(T)$ was determined from experimental lattice parameters,~\cite{Boehmer2018_JPCM30_023001} and $V(0)$ was evaluated by extrapolating the data to the zero-temperature limit. Using $\mathrm{\lambda_{ph-ph}}$ from Eq.~(\ref{eq:linewidth}), the Gr\"{u}neisen parameter $\gamma$, which relates phonon energy and thermal expansion, was obtained from Eq.~(\ref{eq:energy}) as the only remaining free parameter. 

Both the energies and linewidths of the \Algph mode 
are well reproduced for all experimental strain conditions using the same set of parameters (dashed and solid black lines in Fig.~\ref{fig:Figure4}), with $\lambda_{\mathrm{ph\text{-}ph}} \approx 0.25$ and $\gamma \approx 0.4$. This consistency is expected since the \Algph mode corresponds to vibrations of Se atoms along the $c$-axis, and the applied in-plane strain has only a minor effect on the out-of-plane bonding environment. Consequently, the oscillator potential of the phonons remains essentially unchanged. 

The \Blgph mode, corresponding to out-of-phase vibrations of the Fe atoms along the $c$-axis, exhibits a moderate and systematic response to strain. The phonon energy and linewidth evolve smoothly without any anomalous behavior, indicating that this mode is only weakly affected by strain (see Fig.~S5 of the SI). The corresponding phonon–phonon coupling constants and Grüneisen parameters are summarized in Table~1 of the SI. $\gamma\approx 2$ is several times larger than that of the \Algph mode and reflects the stronger coupling of Fe vibrations to lattice contraction and in-plane bonding geometry, as out-of-phase Fe displacements directly perturb the Fe–Fe network.

We now turn to the most crucial and delicate part of the results, the analysis of the additional $\Algprime$ mode and its evolution with strain and temperature. Figure~\ref{fig:Figure4} summarizes the behavior of the \Algph and $\Algprime$ peaks under different strain configurations. 

We first examine the energy offset of the \Algprime mode with respect to the \Algph. In the unstrained sample ($\varepsilon_{\mathrm{zero}}$), the maximal energy difference of about 2~\wn occurs at the lowest temperature where the $\Algprime$ mode is detected. Applying a nematic distortion ($\varepsilon_{\Blg}$) further reduces this offset, whereas $\varepsilon_{\BZg}$ strain produces the opposite trend, driving the $\Algprime$ mode even further to lower energies, with its separation from the \Algph exceeding 3~\wn at the corresponding lowest temperature. In all cases, the $\Algprime$ mode shifts toward the \Algph upon heating, reflecting the reduction of their energy difference. This strain-dependent evolution indicates that the relative position of the $\Algprime$ mode is controlled by the symmetry of the applied strain field and its coupling to the electronic degrees of freedom, an aspect analyzed in detail in the Discussion section.

The temperature dependence of the linewidth of the $\Algprime$ mode is shown in Fig.~\ref{fig:Figure4}~(f–j). For the $\varepsilon_{\mathrm{zero}}$ and $\varepsilon_{\Blg}^{\mathrm{{low, high}}}$ strain configurations, the linewidth closely follows that of the \Algph mode, consistent with a similar anharmonic broadening. In contrast, under $\varepsilon_{\BZg}^{\mathrm{{low, high}}}$, the peak becomes noticeably narrower at lower temperatures than in all other cases. This behavior likely reflects the strain-induced deviation from the natural nematic axis, which enhances local structural disorder, resulting in an apparently sharper line shape.

\subsection*{Discussion}

The emergence of the additional features in the \Alg scattering channel reflects an interplay between lattice vibrations and electronic degrees of freedom in the 11-type iron-based superconductors. In FeS, an overtone excitation resides inside the gap of the PDOS and originates from a two-phonon scattering process involving acoustic branches close to Brillouin zone boundary in the $M$ and $A$ regions [see Fig.~S3 (a) of SI]. Its activation inside this gap has been attributed to enhanced EPC in FeS, where acoustic phonons acquire sufficient coupling strength to produce the overtone near twice their characteristic energy.~\cite{Baum2018_PRB97_054306} As shown in Fig.~S3~(a) of SI, the energies of the acoustic branches at $M$ and $A$ in FeS lie approximately at half of those of the $\Alg^{\mathrm{S,1}}$. Small discrepancies between calculated and experimentally observed phonon energies are within the expected accuracies of the lattice-dynamical models.

However, the situation in \FESES is notably different for small $x$: the PDOS does not have a gap, and an additional feature emerges in the \Alg channel even at the lowest substitution levels ($x \approx 0.04$ – $0.05$)~\cite{PhysRevB.106.094510, zhang_FeSe1-xSx_2021} [see Fig.~S4 of SI]. The structures appearing alongside the allowed phonons are difficult to attribute solely to defect-induced scattering, as the degree of disorder introduced by such a low, isoelectronic and isostructural substitution is unlikely to produce pronounced new modes. Instead, the persistence and systematic evolution of these features favor a scenario where EPC of some acoustic branches is enhanced at the Brillouin zone boundary. The softening and broadening of the \Algph energy (for $x<0.2$) and linewidth,~\cite{PhysRevB.106.094510} respectively, may originate from decay channels becoming available for the \Algph phonon through EPC.

In FeSe, a similar feature labeled as \Algprime is observed, but only within a narrow temperature range around the nemato-structural transition. Unlike in \FESES, where the additional structure develops on the high-energy side of the \Algph line (Fig.~S4) and persists over a broad temperature interval,~\cite{zhang_FeSe1-xSx_2021} in FeSe the additional mode  appears on the low-energy side of the \Algph mode and vanishes once the system stabilizes into either the fully developed nematic phase below \Ts or above the range of strong fluctuations (Fig.~\ref{fig:Figure4}). We interpret the \Algprime mode in terms of an  enhanced EPC near the transition, where fluctuations of both electronic and lattice character are maximal.

The $X$-$R$ direction of the phonon dispersion [2-Fe cell, see Fig.S2 (c)], particularly the $X$ point seems to  play an important role in this behavior. In the nematic phase where the Fe-Fe distance is modulated, the electronic structure at $M$, including the lifting of the $d_{xz}/d_{yz}$ degeneracy, is reconstructed. Then, the formation of the Van Hove singularities and the missing electron pocket, renders the Fermi surface highly anisotropic and enhances the susceptibility of the electronic states to lattice distortions. In this critical region, acoustic phonons near the $X$ point (see Fig.~\ref{fig:Figure2}) become exceptionally susceptible to local symmetry breaking, giving rise to the observed $\Algprime$ mode. This sensitivity also explains the pronounced dependence on both the direction and magnitude of applied strain: when strain couples constructively to the nematic distortion (\Blg strain), the fluctuating lattice potential stabilizes the $\Algprime$ mode within a narrow temperature window; when applied along the diagonal of the 1-Fe cell (\BZg strain) it competes with the natural nematic orientation, broadening the fluctuation regime and modifying the EPC anisotropy.

In crystals detwinned along the Fe-Fe bond direction, the strain aligns with the natural nematic axis and promotes domain ordering. In this configuration, the additional mode gradually merges with the \Algph phonon and persists over a narrower temperature range, consistent with the suppression of critical fluctuations once the lattice becomes more ordered. Conversely, \BZg strain drives the system away from the preferred nematic orientation, introducing competing local distortions and modifying the anisotropy of EPC. As a result, the temperature range over which  the $\Algprime$ mode is observed broadens and the peak itself becomes more sharply defined (Fig.~\ref{fig:Figure2} and ~\ref{fig:Figure4}). This contrast highlights the sensitivity of the FeSe lattice to the direction of applied strain and reveals that the oscillator strength and coherence of the $\Algprime$ mode are governed by the balance between nematic ordering and a modulation of the EPC by lattice distortions.

Thus, while the phonon anomaly in FeS and FeSe may be traced back to the acoustic branches near the zone boundary, the underlying driving mechanisms differ fundamentally. In FeS, the effect reflects static enhancement of EPC within a structurally stable phase, while in FeSe it emerges dynamically from the coupling between critical fluctuations and low-energy lattice modes, amplified by the fluctuating Fermi surface topology near the $M$ point.

These findings contribute to the broader effort to clarify the complex behavior of FeSe, a system in which the interrelation between its structural, magnetic, and electronic degrees of freedom remains elusive. The pronounced, direction-dependent lattice response revealed here demonstrates the exceptional sensitivity of the EPC to weak symmetry-breaking perturbations.

\section{Methods}

The FeSe single crystals were grown using chemical vapor transport (CVT).~\cite{PhysRevB.100.020501} The high-purity sample had a transition from a tetragonal to an orthorhombic structure  at $\Ts=89.1$\,K.~\cite{Baum2019_CP2_14} The measurements in the temperature range between 78 and 300\,K were performed in KONTI CryoVac He-flow cryostat capable of a base vacuum of $10^{-6}$\,mbar. The sample was cleaved immediately before evacuating the cryostat.   

We attached the sample to the actuators of an in-house-made cell (see SI) for applying uniaxial strain, using a two-component epoxy adhesive (Tokyo Measuring Instruments Laboratory EA-2A) developed for usage in cryogenic environment (4 - 323\,K).  A schematic of the way the sample is mounted on the strain cell is shown in Fig.~\ref{fig:Figure1}~(a) and (b). The direction of the strain was either parallel [Fig.~\ref{fig:Figure1}~(a)] or oriented at 45\grd [Fig.~\ref{fig:Figure1}~(b)] with respect to the axes of the 1-Fe unit cell. The deformations are given in terms of the irreducible representations \Blg (orthorhombic deformation of the 1-Fe cell) and  \BZg (rhombohedral deformation of the 1-Fe cell) as described by Ikeda \textit{et al.}~\cite{PhysRevB.98.245133}

The Raman scattering experiment was performed using a Jobin Yvon T64000 triple spectrometer in back-scattering micro-Raman configuration. To achieve a resolution close to 1\,${\rm cm^{-1}}$ per pixel on the CCD detector, we used diffraction gratings with 1800 grooves/mm. In back-scattering configuration, the incident and scattered photons propagate nearly parallel the crystallographic $c$-axis, and the polarizations are in the $ab$ plane as shown in Fig.~\ref{fig:Figure1}~(f). We use Porto notation for describing the light polarizations and assume that $b = a$ above and below \Ts because of the very small orthorhombic distortion and $b\perp a$, $a'=(a+b)/\sqrt2$, $b'=(a-b)/\sqrt2$. Then $(aa)$ means that both incoming and outgoing polarizations \vei and, respectively, \ves are parallel to the $a$ axis.

For excitation we used a mixed-gas Ar$^+$/Kr$^+$ laser (Coherent Innova 70C) operating at the 514.5\,nm emission line.  The laser beam was focused using a microscope objective lens having an ultra-long working distance and $\times$50 magnification, allowing a spot size of approximately 6\,$\mu$m. The power of the incoming polarized light was adjusted for each measurement to deliver 1\,mW on the sample surface or 0.5\,mW absorbed by the material inducing a heating of approximately 4\,K. All temperatures shown here were corrected for this heating. The experimental Raman intensities were divided by the the Bose-Einstein factor yielding response functions $R\chi''(\omega)$. For describing the phonon lines Voigt profiles (a convolution of a Lorentzian and a Gaussian profile) turned out to be superior to simple Lorentzians since the linewidths are very small. The Gaussian and the Lorentzian profiles account for the resolution of the spectrometer (typically 1\,\wn) and the linewidth of the phonon peaks, respectively.

\section*{Acknowledgments}

We thank Andres Baum for valuable discussions regarding the experimental setup, Jonas Bekaert for insightful discussions related to lattice dynamics, Vladimir Damljanović for helpful discussions concerning the symmetry analysis of the system  and Sanja Djurdji\'c Mijin for contributions during the Raman experiment. The authors acknowledge funding provided by the Institute of Physics Belgrade, through a grant from the Ministry of Education, Science and Technological Development of the Republic of Serbia, Project F-134 of the Serbian Academy of Sciences and Arts.  This work was supported by the Science Fund of the Republic of Serbia (SF), PROMIS-2, No. 10925 – DYNAMIQS, and by the SF PROMIS Program, No. 6062656 – StrainedFeSC. This work has received funding from the European Union’s Horzion Europe
research and innovation programme under grant agreement No 101185375 (HIP-2D-QM). The collaboration between the
Institute of Physics Belgrade and the IFW Dresden was supported by the German Academic Exchange Service (DAAD)
through project 57703419 and the Ministry of Science, Technological Development and Innovations of the Republic of Serbia.  C. P. acknowledges financial support from Shanghai Key Laboratory of MFree, China (No. 22dz2260800) and Shanghai Science and Technology Committee, China (No. 22JC1410300). DFT calculations were performed using computational resources at Johannes Kepler University (Linz, Austria). Materials synthesis was supported by the U.S. DOE-BES, Division of Materials Science and Engineering, under Contract DE-SC0012704 (BNL).

\section*{Author contributions statement}

J.B., A.M., R.H., and N.L. designed and set up the experiment. 
A.W. and C.P. prepared the samples. J.P., A.\v{S}., and E.D.L. performed the DFT calculations. 
E.S.B., M.V.M. and Z.V.P. contributed to scientific discussions and interpretation of the results. 
J.B., A.M., T.B., B.V., M.O., N.P., and N.L. carried out the Raman scattering measurements. 
All authors contributed to data analysis, discussion of the results, and writing of the manuscript.



\end{document}